\documentclass{aa}  
\usepackage{graphicx}
\usepackage{txfonts}
\usepackage{natbib}
\usepackage{dcolumn}

\newcolumntype{.}{D{.}{.}{-1}}
\newcolumntype{;}{D{;}{.}{7}}
\bibpunct{(}{)}{;}{a}{}{,} 

\newcommand{\solm}{M$_{\odot}$\ }

\begin{document}

\authorrunning{Eckart et al.}
\titlerunning{Sgr~A* Flares}
\title{Simultaneous NIR/sub-mm observation of flare emission from Sgr~A*}
\subtitle{}

\author{A. Eckart$^{1,2}$,
       R. Sch\"odel$^{3}$,
       M. Garc\'{\i}a-Mar\'{\i}n$^1$,
       G. Witzel$^1$,
       A. Weiss$^2$,
       F. K. Baganoff$^4$,
       M. R. Morris$^5$,
       T. Bertram$^1$,
       M. Dov\v{c}iak$^{6}$, 
       W.J. Duschl$^{7,8}$,  
       V. Karas$^{6}$, 
       S. K\"onig$^{1}$,
       T. P. Krichbaum$^{2}$,
       M. Krips$^{9, 14}$, 
       D. Kunneriath$^{1,2}$,
       R.-S. Lu$^{2,1}$,
       S. Markoff$^{10}$,
       J. Mauerhan$^5$,
       L. Meyer$^5$,
       J. Moultaka$^{11}$,
       K. Mu\v{z}i\'c$^{1}$,
       F. Najarro$^{12}$,
       J.-U. Pott$^{5,13}$,
       K. F. Schuster$^{14}$,
       L. O. Sjouwerman$^{15}$,
       C. Straubmeier$^1$,
       C. Thum$^{14}$,
       S. N. Vogel$^{16}$,
       H. Wiesemeyer$^{17}$,
       M. Zamaninasab$^{1,2}$,
       J. A. Zensus$^{2}$
}

\offprints{A. Eckart (eckart@ph1.uni-koeln.de)}

   \institute{ I.Physikalisches Institut, Universit\"at zu K\"oln,
              Z\"ulpicher Str.77, 50937 K\"oln, Germany
              \email{eckart@ph1.uni-koeln.de}
         \and
             Max-Planck-Institut f\"ur Radioastronomie, 
             Auf dem H\"ugel 69, 53121 Bonn, Germany
         \and
  Instituto de Astrof\'isica de Andaluc\'ia, Camino Bajo de
    Hu\'etor 50, 18008 Granada, Spain 
         \and
 Center for Space Research, Massachusetts Institute of
           Technology, Cambridge, MA~02139-4307, USA 
         \and
  Department of Physics and Astronomy, University of California, 
     Los Angeles, CA 90095-1547, USA
         \and
 Astronomical Institute, Academy of Sciences, 
        Bo\v{c}n\'{i} II, CZ-14131 Prague, Czech Republic 
         \and
 Institut f\"ur Theoretische Physik und Astrophysik,
        Christian-Albrechts-Universit\"at zu Kiel, Leibnizstr. 15
        24118 Kiel, Germany 
         \and
 Steward Observatory, The University of Arizona, 933 N. 
     Cherry Ave. Tucson, AZ 85721, USA
         \and
 Harvard-Smithsonian Center for Astrophysics, SMA project, 
     60 Garden Street, MS 78 Cambridge, MA 02138, USA
         \and
 Astronomical Institute `Anton Pannekoek', 
        University of Amsterdam, Kruislaan 403,
        1098SJ Amsterdam, the Netherlands
         \and
 LATT, Universit\'e de Toulouse, CNRS, 14, Avenue Edouard Belin, 31400 Toulouse, France
         \and
 DAMIR, Instituto de Estructura de la Materia, 
        Consejo Superior de Investigaciones Cient\'ificas, 
        Serrano 121, 28006 Madrid, Spain
         \and
W.M. Keck Observatory (WMKO), CARA, 65-1120 Mamalahoa Hwy., Kamuela, HI-96743, USA
         \and
 Institut de Radio Astronomie Millimetrique, Domaine Universitaire, 
    38406 St. Martin d'Heres, France
         \and
 National Radio Astronomy Observatory,
       PO Box 0, Socorro, NM 87801, USA
         \and
 Department of Astronomy, University of Maryland, College Park, 
    MD 20742-2421, USA
         \and
 IRAM, Avenida Divina Pastora, 7, N\'ucleo Central, 
      E-18012 Granada, Spain
             }

\date{Received  / Accepted }

\abstract{We report on a successful, simultaneous observation and modeling 
of the sub-millimeter to near-infrared flare emission of the Sgr~A* counterpart 
associated with the super-massive (4$\times$10$^6$\solm) black hole at the Galactic center.  
}{
We study and model the physical processes giving rise to the variable emission of Sgr~A*.
}{
Our non-relativistic modeling is based on simultaneous observations that have been  
carried out on 03 June, 2008. We used the NACO adaptive
optics (AO) instrument at the European Southern Observatory's Very Large
Telescope and the LABOCA bolometer at the Atacama Pathfinder Experiment (APEX).
We emphasize the importance of a multi-wavelength simultaneous fitting as a tool 
for imposing adequate constraints on the flare modeling.}{
The observations reveal strong flare activity in the 0.87~mm (345~GHz) sub-mm domain 
and in the 3.8$\mu$/2.2$\mu$m NIR.
Inspection and modeling of the light curves  show that the 
sub-mm follows the NIR emission with a delay of 1.5$\pm$0.5 hours.
We explain the flare emission delay
by an adiabatic expansion of the source components.
The derived physical quantities that describe the flare emission give a source component 
expansion speed of v$_{exp} \sim 0.005$c, source sizes around one Schwarzschild radius
with flux densities of a few Janskys, and 
spectral indices of $\alpha$=0.8 to 1.8, corresponding to particle 
spectral indices $\sim$2.6 to 4.6.
At the start of the flare the spectra of these components peak at frequencies 
of a few THz.
}{
These parameters suggest that the adiabatically expanding source components
either have a bulk motion greater than v$_{exp}$ or the expanding material 
contributes to a corona or disk, confined to the immediate surroundings of Sgr~A*.
}


   \titlerunning{VLT/APEX flare from Sgr~A*}
   \authorrunning{Eckart, Sch\"odel, Garc\'{\i}a-Mar\'{\i}n, Witzel, Weiss  et al.}  
   \maketitle
%

\section{Introduction}
\label{section:Introduction}

\begin{table*}
\centering
{\begin{small}
\begin{tabular}{cccll}
\hline
Telescope & Instrument & $\lambda$ & UT and JD & UT and JD \\
Observing ID & & & Start Time & Stop Time \\
\hline
VLT UT~4       & NACO      & 2.2~$\mu$m  & 2008 3 June 08:37:23 & 3 June 09:58:59 \\
               &           &             & JD 2454620.8593      & JD 2454620.9159 \\
VLT UT~4       & NACO      & 3.8~$\mu$m  & 2008 3 June 04:41:34 & 3 June 8:20:14  \\
               &           &             & JD 2454620.6955      & JD 2454620.8474 \\
APEX           & LABOCA    & 870~$\mu$m  & 2008 3 June 04:56:55 & 3 June 10:22:55  \\
               &           &             & JD 2454620.7062      & JD 2454620.9326 \\
\hline
\end{tabular}
\end{small}}
\caption{Observation log.}
\label{log}
\end{table*}

At the center of the Milky Way, at a distance of only about 8~kpc,
stellar orbits have convincingly proven the existence of a 
super-massive black hole (SMBH) of
mass $\sim$4$\times$10$^6$\solm ~at the position of the
 compact radio, infrared, and X-ray source Sagittarius A* 
(Sgr~A*;
Eckart \& Genzel 1996, 1997,
Eckart et al. 2002,
Sch\"odel et al. 2002,
Eisenhauer et al. 2003,
Ghez et al. 2000, 2005a).
Additional strong evidence of an SMBH
at the position of Sgr~A* comes from the observation of 
rapid flare activity
both in the X-ray and NIR wavelength domains
(Baganoff et al. 2001; Genzel et al. 2003; Ghez et al. 2004, Eckart et al. 2006abc).
Due to its proximity, Sgr~A* provides us with a unique opportunity to
understand the physics and possibly the evolution of SMBHs at the
nuclei of galaxies.  
Variability at radio through sub-millimeter wavelengths has been studied
extensively, showing that variations occur on timescales from hours to
years (e.g.  Mauerhan et al. 2005, Eckart et al. 2006a, 
Yusef-Zadeh et al. 2008, 
Marrone et al. 2008).
Several flares have provided evidence of decaying millimeter and
sub-millimeter emission following simultaneous NIR/X-ray flares.
Although much effort has been invested in monitoring of the galactic 
center variability at these wavelengths, only a handful of simultaneous 
observations aimed at analyzing their correlation have been 
carried out until now (see for instance
Eckart et al. 2006a, 2008ab; Yusef-Zadeh et al. 2006ab, 2007, 2008, Marrone et al. 2008).
 
Here we report on the first measurements of Sgr~A* 
that successfully detected simultaneous flare emission in the near-infrared and sub-millimeter domain
using the ESO VLT and the APEX sub-mm telescopes\footnote{Based on observations with the ESO telescopes at 
the Paranal Observatory under programs IDs:077.B-0028, 79.B-0084, and 81.B-0648.
The sub-mm data are based on observations with the Atacama Pathfinder Experiment (APEX).
APEX is a collaboration between the Max-Planck-Institut f\"ur Radioastronomie,
the European Southern Observatory, and the Onsala Space Observatory}.
In the second section we describe the observations and data reduction
followed by an outline of the flare modeling in section 3.
We conclude with a discussion and summary in sections 4 and 5.

For optically thin synchrotron emission we refer throughout this paper
to photon spectral indices ($\alpha$) using
the convention $S_\nu\propto\nu^{-\alpha}$
and to spectral indices ($p$) of electron power-law distributions
using $N(E)\propto E^{-p}$ with $p=\left(1+2\alpha\right)$.
The assumed distance to Sgr~A* is 8~kpc (Reid 1993), consistent 
with recent results 
(e.g. Lu et al. 2008, Ghez et al. 2005a, 2008, Eisenhauer et al. 2003).

\section{Observations and data reduction}
\label{section:Observations}
The sub-mm regime is of special interest for simultaneous flare measurements, as it 
provides important constraints on the modeling. 
In this general wavelength range synchrotron source components that also radiate 
in the infrared domain become optically thick, and represent the dominant 
reservoir of photons that are then scattered to the X-ray domain through the 
inverse Compton process. 
On 3 June 2008,  substantial progress was made in this matter: as part of a 
larger campaign, Sgr~A* was simultaneously observed in the sub-millimeter and 
NIR wavelength domains using the European Southern Observatory facilities. 
These observations result in the first simultaneous successful detection 
of strongly variable emission in both wavelength domains. 
In the following subsections we describe the data acquisition and 
reduction for the individual telescopes. 
The basic details of the individual observing sessions are given in Table\,\ref{log}.

\begin{table}
\centering
{\begin{small}
\begin{tabular}{ccccccc}
\hline
 $\lambda$ & DIT & NDIT & N & Pixel Scale & Seeing  \\
 & &  & & &  \\
\hline
 2.2\,$\mu$m & 10\,s  & 4   & 80 & 0$\farcs$027    & $\sim$$1\farcs0-1\farcs4$ \\
 3.8\,$\mu$m & 0.2\,s & 150 & 150 & 0$\farcs$027    & $\sim$$1\farcs1-1\farcs7$  \\
\hline
\end{tabular}
\end{small}}
\caption{Details of NIR observations. 
\label{NIRObs}}
\end{table}

\begin{table*}
\begin{center}
\begin{small}
\begin{tabular}{llcccccccc} \hline
flare & spectral& flare start & flare stop & flare peak & FWZP & FWHM & total & flare & offset \\
I.D.  & domain  &  time    & time  & time    & (min)& (min)& peak & peak & \\
      & & & & & & & (mJy) & (mJy) & (mJy) \\ \hline
I     & NIR L'-band & $<$04:43          & 06:20$\pm$10~min. & 05:15$\pm$5~min. & $>$90     & 40$\pm$5   & 32$\pm$2   & $\ge$29   & $\le$3    \\
II    & NIR L'-band & 06:20$\pm$10~min. & 07:20$\pm$10~min. & 06:50$\pm$5~min. & 60$\pm$10 & 40$\pm$5   & 17$\pm$2   & $\ge$14   & $\le$3    \\
III~1 & NIR L'-band & 07:20$\pm$10~min. & $>$8:20           & 08:10$\pm$5~min. & $>$55     & $>$25      & 17$\pm$2   & $\ge$14   & $\le$3    \\
III~2 & NIR Ks-band & $<$8:20           & 08:54$\pm$10~min. & 08:43$\pm$5~min. & $>$20     & 10$\pm$5   & 12$\pm$1   & $\ge$ 6   & $\le$6    \\
IV    & NIR Ks-band & 08:54$\pm$10~min. & 09:48$\pm$10~min. & 09:20$\pm$5~min. & 55$\pm$10 & 22$\pm$5   & 14$\pm$1   & $\ge$ 8   & $\le$6    \\
-     & sub-mm      & $\le$06:00        & 08:50$\pm$20~min. & 07:00$\pm$15~min.& $>$220    & $>$140     & 4250$\pm$200 & 850$\pm$200 & $\sim$3400\\
      & & & & & & & & & \\ 
\hline
\end{tabular}
\end{small}
\end{center}
\caption{
Here we list data of the flares observed by the VLT and APEX on 3 June 2008.
We list the estimated start and stop times, the full width at zero power 
(FWZP) and full width at half maximum (FHWM) values, as well as estimates 
of the flare peaks and light curve minima which 
may be considered as offsets.
The flare peak flux densities are given in 10$^{-3}$ Jansky.
Flux density uncertainties are given as their 1$\sigma$ values.
}
\label{flareprop}
\end{table*}

\subsection{The NIR data}
\label{section:NIR}

Near-infrared (NIR) observations of the Galactic center (GC) were
carried out with the NIR camera CONICA and the adaptive optics (AO)
module NAOS (briefly ``NACO'') at the ESO VLT unit telescope~4 (YEPUN) on
Paranal, Chile, during the night between 2 June and 3 June 2008. In all
observations, the infrared wavefront sensor of NAOS was used to lock
the AO loop on the NIR bright (K-band magnitude $\sim$6.5) supergiant
IRS~7, located about $5.6''$ north of Sgr~A*.  
Details on integration times and approximate seeing during the
observations are listed in Table\,\ref{NIRObs}. 
Here $\lambda$ is the
central wavelength of the broad-band filter used. DIT is the detector
integration time in seconds. NDIT is the number of exposures of
integration time DIT that were averaged on-line by the instrument. N
is the number of images taken. The total integration time amounts to
DIT$\times$NDIT$\times$N. Seeing is the value measured by the
Differential Image Motion Monitor (DIMM) on Paranal at visible
wavelengths. It provides a rough estimate of atmospheric conditions
during the observations.
The atmospheric conditions (and consequently the AO correction) were
stable during the observations.

All observations in the  L'-band (3.8~$\mu$m) and K-band (2.2~$\mu$m)
were dithered to cover a larger area of the GC by mosaic imaging. 
The sky background for the K-band observations was extracted from the median of
stacks of dithered exposures
of a dark cloud -- a region practically empty of stars --
a few arcminutes to the northwest of Sgr~A*.
The L'-band observations were interspersed with frequent off-target exposures 
that served to determine the rapidly varying background in this filter.
All exposures were sky subtracted, flat-fielded, and corrected for
dead or bad pixels. 
Subsequently, PSFs were extracted from these
images with \emph{StarFinder}
(Diolaiti et al. 2000).
The images were deconvolved with the
Lucy-Richardson (LR) algorithm. Beam
restoration was carried out
with a Gaussian beam of FWHM corresponding to the respective
wavelength. The final resolution at 2.2 and 3.8 \,$\mu$m
is  about 60 and 104~milli-arcseconds (mas), respectively.

The flux densities of the sources were measured by aperture photometry with
circular apertures of 52~mas radius and corrected
for extinction, using $A_{K} = 2.8$, and  $A_{L'} = 1.8$.
The relative flux density calibration was carried out using
known K- and L'-band flux densities of IRS16SW, IRS16C, IRS16NE, and IRS21 by 
Blum, Sellgren \& Depoy (1996).
This results in K- and L'-band flux densities of the high velocity star
S2 of 22$\pm$1~mJy and 9$\pm$1~mJy, respectively, which 
compare well with magnitudes and fluxes for S2 quoted by 
Ghez et al. (2005b) and Genzel et al. (2003).
The relative photometry for Sgr~A*
was done with the known fluxes and positions of
9 sources within 1\farcs6 of Sgr~A*. 
The measurement uncertainties for Sgr~A* were obtained on the reference star S2.
The background flux in the immediate vicinity of Sgr~A* was 
obtained by averaging the measurements at six random locations 
in a field located about 0\farcs6 west of Sgr~A* that is
free of obvious stellar sources.

\begin{figure}
\centering
\includegraphics[width=8cm,angle=-00]{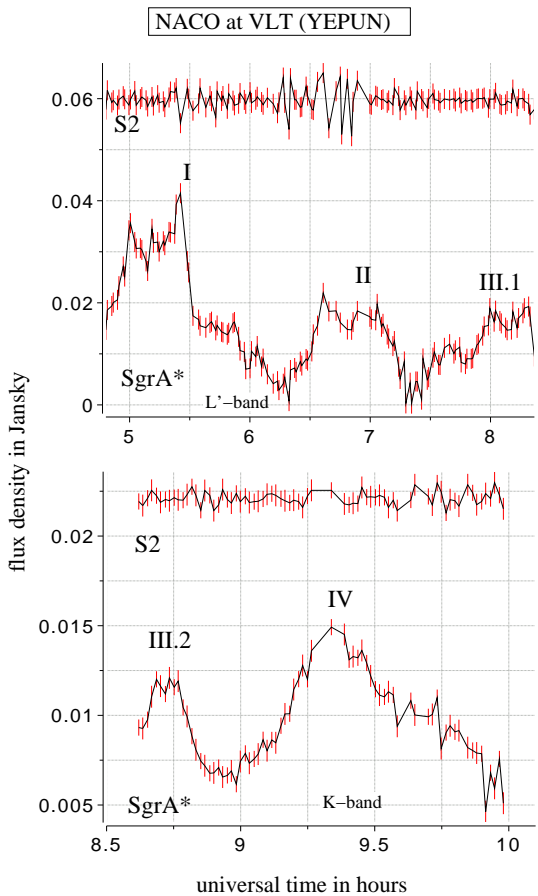}
\caption{\small
The infrared light curve of Sagittarius A* on 3 June 2008. 
The data are represented by vertical red bars ($\pm$1$\sigma$) 
and a black connection line between them. 
Upper panel: L'-band flare events.
For a better representation we added a constant value of 50~mJy to the light curve for the reference star.
Lower panel: The K-band flare emission immediately following the L`-band observations.
In this case no offset was added to the light curve for the reference star.
}
\label{fig:1}    
\end{figure}

The resulting light curves of Sgr~A* and S2 (as a constant reference star) are shown in Fig.\,\ref{fig:1}.  

\vspace{2mm}
\noindent

\vspace{2mm}
\noindent
\begin{figure*}
\centering
\includegraphics[width=18cm,angle=00]{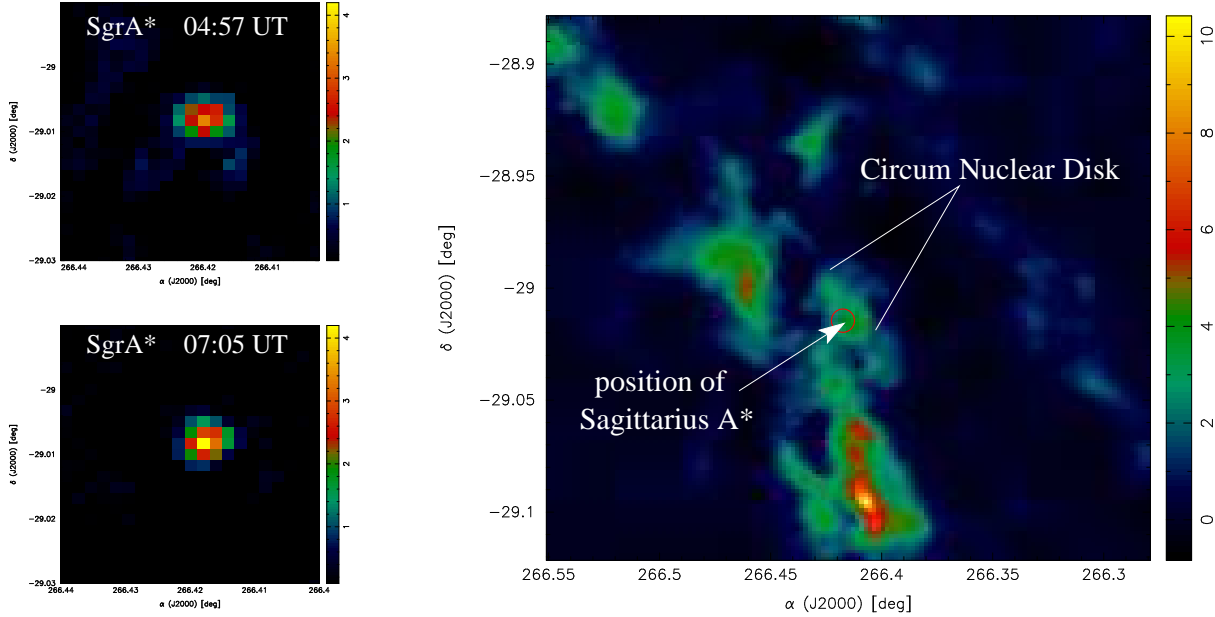}
\caption{\small
Left: Sagittarius A* at the beginning and the peak 
 of the APEX measurements. 
Right: A section of the larger map of the
Galactic center. The location of Sagittarius A* is indicated by a circle.
Sagittarius A* has been subtracted out as this map is used as a reference for
the extended structure. 
The colors code the flux density in Janskys.
}
\label{fig:2}    
\end{figure*}

\subsection{The sub-mm data}
\label{section:submm}
The Galactic center 870 $\mu $m data were taken with the newly commissioned LABOCA bolometer 
array, located on the Atacama Pathfinder EXperiment (APEX) telescope at the Llano Chajnantor, 
Chile, at an altitude of 5105~m.  
The observations were carried out as part of  global multi-wavelength monitoring program in May/June 2008.
The radiation is collected by the APEX telescope and fed through conical horns 
to the detectors of the Large APEX Bolometer Camera (LABOCA), an array of 295 composite 
bolometers that are extremely sensitive to continuum radiation.  
With a total bandwidth of about 60~GHz 
the system is optimized for the 345~GHz atmospheric window.

The full width half maximum (FWHM) of the point spread function 
(PSF) at 870~$\mu$m is $\sim19^{\prime \prime }$. 
To allow for the best possible reconstruction of the 
complex sub-mm emission in the GC region we used on the fly 
(OTF) maps perpendicular to the galactic plane
(Weiss et al. 2008, Siringo et al. 2007).
The maps were observed with a scanning speed of 3'/s modulating the
source signals even for extended structures into the 
useful post-detection frequency band of LABOCA (0.1-12.5~Hz).
To avoid scanning artefacts in the reconstruction of the
extended thermal emission surrounding Sgr~A*, we derived and averaged 
maps from the data with an inclination of 
-10$^{\circ}$, 0$^{\circ}$ and +10$^{\circ}$ with 
respect to an axis orthogonal to the galactic plane.
The mapping steps perpendicular
to the scanning direction was 30$^{\prime \prime }$ and the integration
time 280~seconds, yielding an rms noise level
of $\sim$150~mJy/beam for each map. 

Immediately after each galactic center map, either G10.62 or 
IRAS16293-2422 were observed as secondary calibrator sources. 
We reduced the data with the BoA\footnote{BoA: 
http://www.astro.uni-bonn.de/boawiki/Boa} software package. 
The data reduction process included correction for atmospheric zenith 
opacity (towards the Galactic center $\tau$$\sim$0.3 up to 0.7), 
flat-fielding, de-spiking, correlated sky noise removal, 
and the removal of further 
correlated noise due to instrumental effects.

After correcting for pointing
offsets between individual maps (determined from the
position of Sgr~A* in each map) all scans were co-added
to obtain high S/N, 48`$\times$25` fully sampled maps
of the sub-mm emission in the GC regions. From this combined map 
the point source Sgr~A* was modeled with a Gaussian and taken out.
The uncertainty in that process will result in a small offset of the 
sub-mm light curve. We estimate that this effect is less than 10\% 
of the peak flux density of SgrA*.
In this way we created a model image of the extended 
870$\mu$m emission surrounding Sgr~A*
with Sgr~A* effectively set in an off state (see Fig.\,\ref{fig:2}). 

Each data point of the Sgr~A* sub-mm light curve (see Fig.\,\ref{fig:3}) 
was derived from the model subtracted maps, modeling a Gaussian source 
and deriving the peak. 
In order to provide a stable and consistent calibration 
we used the $\tau$ values derived 
from the sky dip and radiometer measurements, as well as 
secondary flux calibrators and a comparison to the reference map.
Between the individual observing epochs 
the flux densities of the secondary flux calibrators
vary by $\le$10\%.
From a comparison of different reference sources, we
estimate the relative point-to-point uncertainty in the calibration of the light curve 
measurements to be of the order of 4\% (1~$\sigma$).
The remaining flux density variations in the light curve in Fig.\,\ref{fig:3}
may arise from residual fast and uncompensated opacity or pointing changes.

\section{Flare Analysis}
\label{section:FlareAnalysis}

\subsection{Adiabatically expanding source components}
\label{section:Adiabatic}
Our basic assumption to model the sub-mm light curves is the presence of an expanding 
uniform blob of relativistic electrons with an energy spectrum $n(E) \propto E^{-p}$ 
threaded by a magnetic field. As a consequence of the blob expansion, 
the magnetic field inside the blob declines as $R^{-2}$, the energy of relativistic 
particles as $R^{-1}$ and the density of particles as $R^{-3}$
(van der Laan 1966).  
The synchrotron optical depth 
at frequency $\nu$ then scales as 
\begin{equation}
    \tau_\nu = \tau_0 \left(\frac{\nu}{\nu_0}\right)^{-(p+4)/2}
    \left(\frac{R}{R_0}\right)^{-(2p+3)}
    \label{eq:taunu}
\end{equation}
and the flux density scales as
\begin{equation}
    S_\nu = S_0 \left(\frac{\nu}{\nu_0}\right)^{5/2} 
    \left(\frac{R}{R_0}\right)^3 
    \frac{1-\exp(-\tau_\nu)}{1-\exp(-\tau_0)}
    \label{eq:Snu}
\end{equation}
Here R$_0$, S$_0$, and $\tau_0$ are the size, flux density, and
optical depth at the peak frequency of the synchrotron spectrum $\nu_0$.
The goal of the present model is to combine the description of an adiabatically 
expanding cloud with a synchrotron self-Compton formalism, as this is the most 
likely physical scenario to explain the delay between the sub-mm and the 
simultaneous near-IR and X-ray peaks. Thus, we use the definition of
$\tau_0$ as the optical depth corresponding to the frequency at which the 
flux density is a maximum (van der Laan 1966), rather than the definition of
$\tau_0$ as the optical depth at which the flux density for any
particular frequency peaks (Yusef-Zadeh et al.\ 2006b). 
This implies that $\tau_0$ depends only on $p$ through the condition

\begin{equation}
    e^{\tau_0} - \tau_0(p + 4)/5 - 1 = 0
    \label{eq:tau0}
\end{equation}

\noindent
and ranges from, e.g., 0 to 0.65 as $p$ ranges from 1 to 3.
Therefore, given the particle energy spectral index $p$ and
the peak flux $S_0$ in the light curve at some frequency $\nu_0$, this
model predicts the variation in flux density at any other frequency
as a function of the expansion factor $(R/R_0)$.

Finally, a model for $R(t)$ is required to convert the dependence on
radius to time: we adopt a simple linear expansion at
constant expansion speed v$_{exp}$, so that $R-R_0 = $v$_{exp}\,(t-t_0)$.
For $t \le t_0$ we have made the assumption that the source has an optical 
depth that equals its frequency dependent initial value $\tau_{\nu}$ at $R = R_0$.
So in the optically thin part of the source spectrum the flux initially
increases with the source size at a constant $\tau_{\nu}$ and then
decreases due to the decreasing optical depth as a consequence of the expansion.
For the $\sim$4$\times$10$^6$\solm ~super-massive black hole at 
the position of Sgr~A*,
one Schwarzschild radius is R$_s$=2GM/c$^2$$\sim$10$^{10}$~$m$
and the velocity of light corresponds to 
about 100~R$_s$ per hour. 
For $t > t_0$ the decaying flank 
of the curve can be shifted towards later times by first,
increasing the turnover frequency $\nu_0$ or the initial source size $R_0$,
and second, by lowering the spectral index $\alpha_{synch}$ or the peak 
flux density $S_0$.
Increasing the adiabatic expansion velocity v$_{exp}$ shifts the peak of
the light curve to earlier times.
Adiabatic expansion will also result in a slower decay rate and
a longer flare timescale at lower frequencies.

\vspace{2mm}
\noindent
\begin{figure}
\centering
\includegraphics[width=8cm,angle=-00]{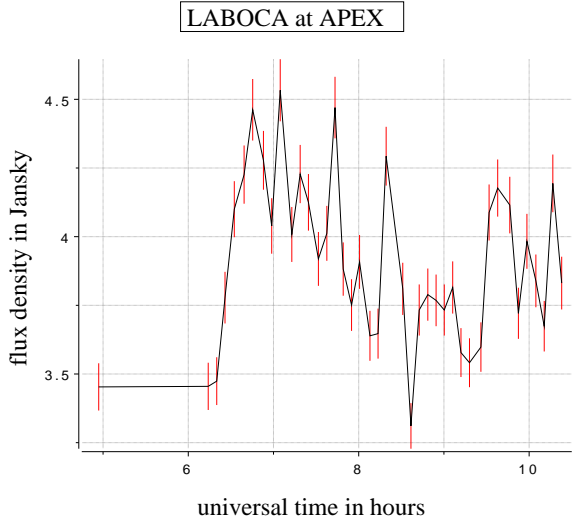}
\caption{\small
The sub-mm light curve of Sagittarius A* on 3 June 2008.
The data are represented by vertical red bars ($\pm$1$\sigma$ in length) 
and a black connection line between them.
}
\label{fig:3}    
\end{figure}

\vspace{2mm}
\noindent
\begin{figure}
\begin{center}
\includegraphics[width=8cm,angle=-00]{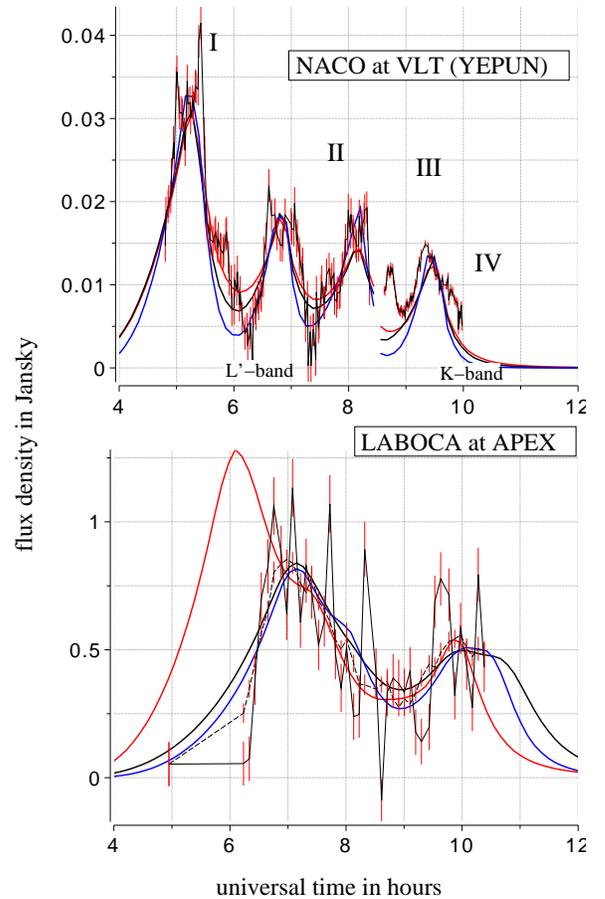}
\caption{\small
The infrared (top) and sub-mm (bottom) light curve of Sagittarius A*.
The data are represented by vertical red bars ($\pm$1$\sigma$) and a black connection line 
between them. 
The dashed line represents a smoothed version of the data 
(after application of a 7 point sliding average for all data points 
exept the first).
The models as described in the text and in Tab.~\ref{modeldata}
(blue, black, red for models A, B, C, respectively) 
are shown as solid lines 
(blue: $\sim$2.5~THz;
black: $\sim$2~THz;
red: $\sim$1~THz cutoff frequency in the source component spectra.)
}
\label{fig:4}    
\end{center}
\end{figure}

\subsection{Modeling the flare emission}
\label{subsection:Modeling}
The combined L'- and K-band data show violently variable emission with at least 
4 prominent flare events (I-IV, see Fig.\,\ref{fig:1}). 
As we switched from L- to K-band during an ongoing flare we attribute the 
L-band flare section III.1 and the K-band flare section III.2 to the same
NIR flare III with a duration from about 7:30~UT to about 9:00~UT.

As can be seen from the first data points in Fig.\,\ref{fig:3},
the sub-mm data start with a low flux density at the time of the first 
NIR flare (Fig.\,\ref{fig:1}). 
The gap shown in the sub-mm data between 5~h and 6~h 
occurred during culmination of Sgr~A*, 
when the source rises above the elevation limit (80$^\circ$) 
for observations with APEX. 
In the sub-mm domain the variable emission is dominated by a single flare 
that is wider than the individual NIR flares
(see Table~\ref{flareprop}).
The so far six reported coordinated SgrA* measurements
that include sub-mm data 
(Eckart et al. 2006b; Yusef-Zadeh et al. 2006b; Marrone et al. 2008, including this work) have
shown that the observed submillimeter flares follow the largest event
observed at the shorter wavelengths
(NIR/X-ray; see detailed discussion in Marrone et al. 2008).
We therefore assume that the sub-millimeter 
flare presented here is related to the observed IR flare events.

{\it Separating the variable sub-mm emission:}
In order to separate the rapid intra-day flare emission from the
sub-mm emission that varies much slower on time scales of days,
a constant flux density level has been subtracted.
In most instances Sgr~A* has been tracked over long 
time spans on each day to obtain light curves.
We have determined the median and median deviation of the lowest
flux density measurements of the obtained light curves. 
This allows us to determine
a flux density contribution that is constant or only slowly variable on 
time scales much longer than a single day.
For this purpose we compared 5 APEX light curves taken 
with LABOCA between May 27 and June 3 (centered at a wavelength of 
0.87~mm, Garc\'{\i}a-Mar\'{\i}n 2008, in prep.) 
as well as 6 CSO measurements at 0.85~mm wavelength by Yusef-Zadeh et al. (2006a, 2008)
and 2 SMA light curves at 1.3~mm wavelength by Marrone et al. (2008).
The SMA interferometer resolves out extended emission on scales of a few arcesonds. 
In the case of the APEX measurements a model of
the extended emission on scales of 19'' is subtracted.
The flux density offsets of the different light curves are very similar. 
Therefore we attribute them to be intrinsic to the emission of Sgr~A*.
The comparison results in a constant offset value of 3.0$\pm$0.5~Jy.
For the the lowest submillimeter flux densities obtained on June 3 
we find a constant contribution of 3.4~Jy.
This value is in very good agreement with the median derived above.
Therefore we subtracted it from the light curve in order
to extract the intra-day variable part of the sub-mm data.

{\it Adiabatic expansion modeling:}
We model the 4 flare events I - IV and the smoothed version of 
the sub-millimeter data with 4 adiabatically expanding
source components (see dashed line in Fig.\,\ref{fig:4}.  
The calculations show that the initial synchrotron 
self-absorption cutoff frequency of the source components is a 
critical parameter that allows us to distinguish between separate models.
From our modeling we can constrain 
the cutoff frequency between 1 and 2.5 THz 
(model A:2.5~THz, model B:2~THz and model C: 1~THz),
and establish from inspection and modeling of the light curves 
a delay between the sub-mm and NIR emission  of 1.5$\pm$0.5 hours.
Model fits to the variable sub-mm emission are 
shown in comparison to the data in Fig.\,\ref{fig:4}.  
The corresponding model parameters are given in Tab.~\ref{modeldata}.
To select the intra-day variable part of the sub-mm light curve 
a flux density of 3.4~Jy has been subtracted from the sub-mm data in Fig.\,\ref{fig:4}.
This amount is attributed to more extended (many Schwarzschild radii) 
source components.
We attribute the time difference between the NIR and sub-mm flares 
to an adiabatic expansion of synchrotron source components with an 
expansion speed of about 0.5\% of the speed of light (1500 km/s). 
While the sub-mm light curve is modeled by the evolving optically thick part
of the spectra, the NIR light curve is modeled through 
the optically thin part of the spectra.
We find that the flares are associated with 
source components that have initial sizes of the order of one Schwarzschild 
radius and spectra that peak around 2~THz. The flux densities are 
a few Janskys and the spectral indices are optically thin between 
the sub-mm and the infrared with 0.6$\le$$\alpha$$\le$1.5. 
Models with significantly different expansion speeds or source sizes 
fail to represent either the extent or shape of the observed flare features.

\begin{figure}
\centering
\includegraphics[width=8cm,angle=-00]{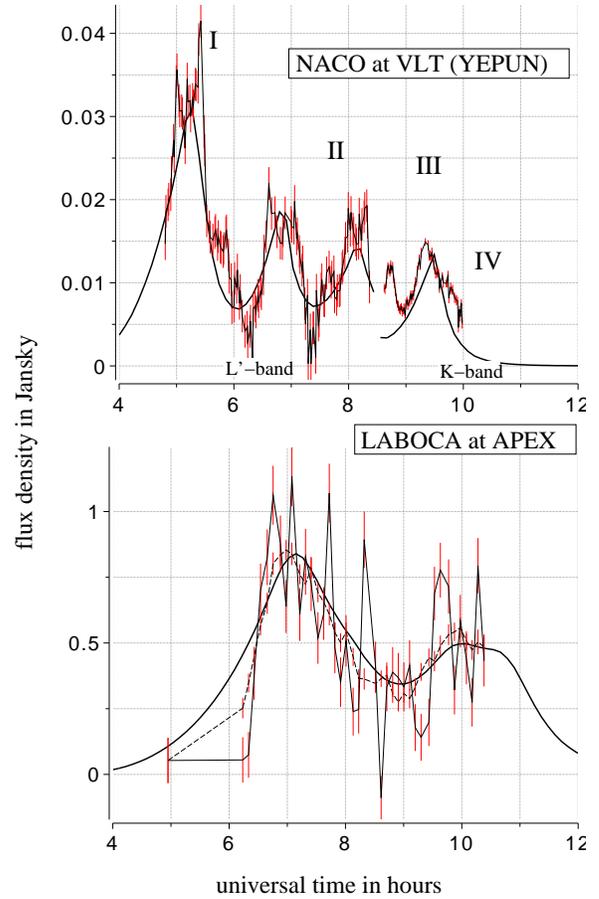}
\caption{\small
Our favored model B compared to the 
infrared (top) and sub-mm (bottom) light curve of Sagittarius A*.
The data are represented by vertical red bars ($\pm$1$\sigma$) and a black connection line 
between them. 
For furher explanation see caption of Fig.\,\ref{fig:4}.  
}
\label{fig:5}    
\end{figure}

Models A and C represent border cases: For higher cutoff frequencies 
than the one used in model A, the required flare fluxes at the THz 
turnover frequencies become increasingly higher to account for the
sub-mm emission after adiabatic expansion. 
At the same time, the spectral index becomes increasingly
steeper in order to meet the NIR flux densities. 
Towards cutoff frequencies well above 2.5~THz the spectral index will 
become steeper than what has been measured in the NIR
(Eisenhauer et al. 2005, Hornstein et al. 2006, 
Gillessen et al. 2006, Krabbe et al. 2006).
On the other hand the mean spectral index for the component in model C 
is closer to the
value of 0.6$\pm$0.2 
obtained by Hornstein et al. (2007), but the predicted source 
component labeled  $I$ shows 
already an unacceptable discrepancy between the flux density observed 
with APEX at 04:56~UT.
This discrepancy increases with decreasing cutoff frequency.
Our favored model B is presented in Fig.\,\ref{fig:5}. 
It has an intermediate cutoff frequency, shows an overlap in parameters 
with models A and C, and consistently 
reproduces the flares in the sub-mm and NIR wavelength domains. 
With a deviation of about 3-4$\sigma$, the agreement of the spectral indicies
with results by Hornstein et al. (2007) lies between that of model A and C.
The spectral indices are in good agreement with that obtained for $\alpha_{NIR/X-ray} \sim 1.3$
by Eckart et al. (2004, 2006a).
This is consistent with the fact that the optically thin synchrotron
spectral index (sub-mm to NIR) is expected to equal the broad band spectral
index of the SSC spectrum.

\vspace{2mm}
\noindent

{\it SSC modeling as an additional constraint:}
We use a synchrotron model with an optically thin spectral index 
$\alpha$ and relativistic electrons with $\gamma_e \sim 10^3$
following Gould (1979) and Marscher (1983).
In the context of Sgr~A* this model has been described by 
Eckart et al. (2004, 2008a).
This formalism allows us to estimate the Synchrotron Self Compton (SSC) X-ray flux densities
and to compare them with values that have been measured before.
In addition it allows us to estimate the  SSC contribution of the
NIR flux density and the magnetic field. 
In our adiabatic expansion modeling we have assumed that 
the NIR flux density is dominated by the synchrotron emission 
of the THz peak expanding synchrotron component.
A strong excess in SSC NIR emission would violate that assumption and would
require modeling the (unfortunately) unknown X-ray flux density as well.
The magnetic field can be obtained via
$B \sim \theta^4 \nu_m^5 S_m^{-2}$
as a direct result of the synchrotron source parameters.
Previous models of the Sgr~A* emission have resulted in 
magnetic field strengths of a few up to 70~Gauss 
(e.g. Eckart et al. 2006a, 2008a, Yusef-Zadeh et al. 2008, Marrone et al. 2008).
The source components assumed for the adiabatic expansion
model should not violate these results.

Our SSC calculations, that reveal typically observed X-ray flux densities 
in the range of a few 10 to 100~nJy with magnetic fields of a few 10~G,
indicate that the best solution lies between models B and C in Tab.~\ref{modeldata}.
For model A we find that the SSC contribution of the NIR flux density
becomes comparable to the synchrotron contribution.
A correspondingly lower predicted NIR synchrotron contribution 
would require even steeper spectral indices. 
The SSC calculations, however, already indicate that the spectral indices 
and cutoff frequencies in model A need to be lower by 
$\Delta$$\alpha$$\sim$0.2 and $\Delta$$\nu$ of a few 100~GHz 
in order to give reasonable solutions that agree with typically 
observed and derived X-ray flux densities and magnetic fields.
Therefore model A clearly appears less preferable.

{\it Quality of the fit:}
Sub-structures deviate by up to 4 times the measurement
uncertainties from the overall structure described by flares I to IV. 
In order to estimate the 
fit quality we scale the assumed uncertainties to within which 
we can model the NIR data to 4$\sigma$.
To compare the NIR and APEX data with the adiabatic expansion models,
we calculated reduced $\chi^2$ values.
Models with a large number of flare or source components would 
increase the number of free parameters and 
allow us to obtain better reduced $\chi^2$ values for the fits
However, more complex models are currently difficult to justify.
Following the ratio of the 41 APEX to 208 NIR data points,
we increased the weight of the APEX data by a factor of 4.
With a total of 4 components 
(each with a time t$_0$, peak flux S$_0$, size R$_0$, and
spectral index), a common cutoff frequency, sub-mm flux offset
and expansion velocity v$_{exp}$ 
we obtain a total of 19 free parameters.
As shown in the Figs.\,\ref{fig:1} to \,\ref{fig:5} the L- and K-band 
data were compared to the L- and K-band model predictions, respectively.
For our favorite model we calculated the uncertainties of the
model parameters by varying them until the reduced $\chi$-values increase by 
$\Delta\chi$$\sim$0.5.
This corresponds to an increase across either of the 4 individual 
flare intervals in the sub-mm or the NIR domain of
$\Delta\chi$$\sim$1.0, corresponding to a confidence 
level of about 68.5\% (i.e. a 1~$\sigma$ uncertainty) for the model 
parameters for each of the flare components.

\begin{table*}
\centering
{\begin{small}
\begin{tabular}{lllcrcrcrrrrrrr}
\hline
model & flare & t$_0$     &v$_{exp}$ &S$_{max, obs}$&$\alpha_{synch}$&R$_0$&$\nu_{0}$ & reduced \\
label & label &    hours  & in $c$   &[Jy]          &                &     & [THz]& $\chi$$^2$ value \\
 \hline
      &       &           &          &              &                &     &       & \\
   A  &  I    &  5.3$\pm$0.2  & 0.0040$\pm$0.0020  &11.4$\pm$3.6  & 1.77$\pm$0.20  & 1.1$\pm$0.10 & 2.50$\pm$0.45& 2.64 \\ 
      &  II   &  6.9$\pm$0.5  &         "          & 4.8$\pm$3.5  & 1.67$\pm$0.10  & 0.7$\pm$0.40 &        "     &   \\    
      &  III  &  8.2$\pm$0.4  &         "          & 5.6$\pm$3.0  & 1.77$\pm$0.10  & 1.0$\pm$0.70 &        "     &   \\
      &  IV   &  9.5$\pm$0.2  &         "          & 4.1$\pm$2.0  & 1.45$\pm$0.10  & 0.6$\pm$0.50 &        "     &   \\
      &       &           &          &              &                &     &      &  \\
   B  &  I    &  5.3$\pm$0.2  & 0.0045$\pm$0.0020  & 7.8$\pm$1.5  & 1.57$\pm$0.10  & 1.3$\pm$0.30 & 2.00$\pm$0.45 & 1.82 \\  
      &  II   &  6.9$\pm$0.5  & "                  & 1.4$\pm$1.5  & 1.25$\pm$0.10  & 0.9$\pm$0.60 &      " & \\ 
      &  III  &  8.2$\pm$0.4  & "                  & 3.3$\pm$1.6  & 1.56$\pm$0.10  & 1.2$\pm$0.70 &      " & \\
      &  IV   &  9.4$\pm$0.2  & "                  & 2.8$\pm$0.5  & 1.33$\pm$0.10  & 0.9$\pm$0.50 &      " & \\
      &       &           &          &              &                &     &       \\
   C  &  I    &  5.3$\pm$0.2  & 0.0053$\pm$0.0020  & 4.0$\pm$1.3 & 1.15$\pm$0.30    & 1.3$\pm$0.60 & 1.00$\pm$0.30 & 2.32 \\ 
      &  II   &  6.9$\pm$0.5  &         "          & 1.3$\pm$1.0 & 1.05$\pm$0.20    & 1.0$\pm$0.50 &      "        & \\
      &  III  &  8.2$\pm$0.5  &         "          & 0.5$\pm$1.3 & 0.85$\pm$0.20    & 1.2$\pm$0.30 &      "        & \\
      &  IV   &  9.4$\pm$0.2  &         "          & 1.4$\pm$0.3 & 1.00$\pm$0.20    & 0.7$\pm$0.40 &      "        & \\
      &       &           &          &             &                &     &       \\
 \hline 
\end{tabular}
\end{small}}
\caption{
Source component parameters for the adiabatic expansion model of the 3 June, 2008 flare.
The flare times t$_0$ are given with respect to the peak of the brighter NIR flares labeled I, II, III, IV.
For the adiabatic expansion velocity v$_{exp}$, the optically thin spectral index $\alpha$$_{synch}$ and
the cutoff frequency $\nu$$_0$ the uncertainties were derived over the entire data set.
In addition to v$_{exp}$ the R$_0$ values are responsible for the 
position and width of the infrared flares peaks in time. 
Different values for $\alpha_{synch}$ are requred to match 
the infrared flux densities.
The derivation of uncertainties and reduced $\chi$$^2$ values 
is described in the text. Model B is the preferred model (see text).
\label{modeldata}}
\end{table*}

\subsection{Discussing the modeling results}
\label{section:Modelingresults}

The adiabatic expansion results in a time difference between the peaks 
in the VLT and APEX light curves of about 1.5$\pm$0.5 hours,
and compare well with the values obtained in 
a global, multi-wavelength observing campaign by our team in 2007.
Back then, two bright NIR flares were traced by 
CARMA (Combined Array for Research in mm-wave Astronomy; 100~GHz) in the US, 
ATCA (Australia Telescope Compact Array; 86~GHz) in Australia,
and the MAMBO bolometer at the IRAM 30m in Spain (230~GHz;
first results given by Kunneriath et al. 2008, Eckart et al. 2008b).
This light curve complements our parallel 13, 7, and 3~mm VLBA run (Lu et al. 2008).
Other recent simultaneous multi-wavelength observations also indicate 
the presence of adiabatically expanding source components with 
a delay between the X-ray and sub-mm flares of about 100 minutes
(Eckart et al. 2006a, Yusef-Zadeh et al. 2008, Marrone et al. 2008).
As pointed out by Eckart et al. (2008ab), a combination of a 
temporary accretion disk with a short 
jet can explain most of the properties associated with infrared/X-ray 
Sgr~A* light curves (Eckart et al. 2008ab).
\\
\\
{\it The adiabatic expansion model:}
The May 2007 polarimetric NIR measurements (Eckart et al. 2008a)
showed a flare event with the highest sub-flare contrast observed until now. 
In the relativistic disk model these data 
provide evidence of a spot expansion and its shearing
due to differential rotation.
An expansion by only 30\% will lower the Synchrotron-Self-Compton 
(SSC) X-ray flux significantly.
In the framework of the spot model this flare event provides additional 
support for expansion of individual source components.
\\
\\
{\it The expansion speed:}
The rapid decay of the observed NIR/X-ray flares 
(e.g. Baganoff et al. 2001; Genzel et al. 2003; Ghez et al. 2004, Eckart et al. 2006abc),
as well as the current results from 
coordinated observing campaigns including sub-mm monitoring 
(Eckart et al. 2006b; Yusef-Zadeh et al. 2006b; Marrone et al. 2008 and this work) 
suggest that non-radiative cooling processes, such as adiabatic expansion, are essential,
although the adiabatic cooling model results in very low expansion speeds.
From modeling the mm-radio flares Yusef-Zadeh et al. (2008)
invoke expansion velocities in the range from $v_{exp}$=0.003-0.1c.
This compares well with the expansion velocity of the order of
0.0045$\pm$0.0020~c that we obtain in our case with the June 2008 data
(see Table~\ref{modeldata}).
These velocities are low compared to the expected 
relativistic sound speed in orbital velocity in the vicinity of the SMBH.
The low expansion velocities 
suggest that the expanding gas can not escape from Sgr~A*
or must have a large bulk motion (see discussions in Marrone et al. 2008 and Yusef-Zadeh et al. 2008).
Therefore the adiabatically expanding source components
either have a bulk motion larger than v$_{exp}$ or the expanding material 
contributes to a corona or disk, confined to the immediate 
surroundings of Sgr~A*.
An expansion of source components through shearing due to differential
rotation within the accretion disk may explain the low expansion velocities.
The recent theoretical approach of hot spot evolution due to shearing is
highlighted in Eckart et al. (2008a) and Zamaninasab et al. (2008; see also Pech\'a\v{c}ek et al., 2008)
\\
\\
{\it Structure of the light curve:}
The flux density variations of Sgr~A* can be explained 
in a disk or jet model (see e.g. discussion in Eckart et al. 2006ab, 2008a), or 
they could be seen as a consequence of an underlying physical process that can 
mathematically be described as red-noise
(Do et al. 2008, Meyer et al. 2008).
In our case we find a light curve structure that consists of maxima separated 
by about 70 to 110 minutes with additional fluctuations of smaller amplitude.
Assuming the presence of a disk and by simultaneous fitting of the 
previously obtained light curve fluctuations and the time-variable 
polarization angle, we have shown that the data can be successfully 
modeled with a simple relativistic hot spot/ring model
(Meyer et al. 2006ab, 2007, Eckart et al. 2006ab, 2008ab).
In this model the broad near-infrared flares ($\sim$100 minutes duration)
of Sgr~A* are due to a sound wave that travels around the SMBH once. 
The sub-flares, superimposed on the broad flare, are thought to be due 
evolving (i.e. expanding) hot-spots that may be relativistically 
orbiting the central SMBH.
The spot emission would then be due to transiently heated and accelerated 
electrons which can be modeled as a plasma component
(scenarios in which spiral wave structures contribute to the observed 
variability are also under discussion, e.g. Karas et al. 2007).

In case of a jet
(see e.g.  Markoff, Bower \&  Falcke 2007, Markoff, Nowak \& Wilms 2005),
the observed flux density variations may more 
likely be a result of the variations in the accretion process 
(or jet instabilities) - possibly followed by an adiabatic expansion 
of the jet components - rather than 
being a result of a modulation from an orbiting spot. 
In this case one may expect that red-noise 
variations on these short times scales are a natural extension 
of the variability found for longer periods.


\section{Summary and discussion}
We have presented new simultaneous measurements of the near-IR sub-mm 
flare emission of the Sgr~A* counterpart associated with the SMBH at the 
Galactic center. The data were obtained in a global campaign carried 
out on 3 June 2008, using NACO at VLT and LABOCA at APEX. 

The highly variable near-IR emission presents four major events 
(I - IV), whose peaks are separated by about $\sim$80~min. 
The peak of the sub-mm light curve
is delayed with respect to that of the near-IR data. 
We argue that this delay is due to the adiabatic expansion 
of synchrotron source components that become optically thin. 
The expansion velocity is about 1500 km/s (0.5\% of the speed of light). 
We model the four flare events with 4 source components. 
In this case the sub-mm light curves of the individual components
are blended, with a time delay for the peaks of about 1.5$\pm$0.5 hours 
with respect to the near-IR data. 

The light curve structure is consistent with
the previously found variability. 
It could be interpreted
as emission from material in relativistic orbits around the SMBH. 
However, variable emission from a jet or an explanation as 
a dominant flux density contribution from a red-noise process 
cannot be excluded as well.

These data show that the VLT/APEX combination is especially well suited 
for very long simultaneous light curves between the NIR and the sub-mm domain. 
Further simultaneous radio/sub-mm data, NIR K- and L-band measurements 
in combination with X-ray 
observations should lead to a set of light curves that will allow us
to prove the proposed model and to discriminate between the individual 
higher and lower energy flare events.
Here Chandra's high angular resolution is ideally suited to separate 
for weak flares the thermal non-variable bremsstrahlung 
and the non-thermal variable part of the Sgr~A* X-ray flux density.

\vspace{0.5cm}
{\bf Acknowledgments:}
{\small
We are grateful to all the ESO PARANAL and Sequitor staff, and especially 
to the members  of the NAOS/CONICA, VLTI, and APEX team.
The observations were made possible through a special effort by the
APEX/ONSALA staff to have the LABOCA bolometer ready for triggering.
Leo Meyer is supported by the DAAD exchange program.
Macarena Garc\'{\i}a-Mar\'{\i}n is supported by the German federal department for 
education and research (BMBF) under the project numbers: 50OS0502 \& 50OS0801.
The X-ray work was supported by NASA through Chandra award G05-6093X.
M. Zamaninasab, D. Kunneriath, and R.-S. Lu,
 are members of the International Max Planck Research School (IMPRS) for 
Astronomy and Astrophysics at the MPIfR and the Universities of 
Bonn and Cologne. R. Sch\"odel acknowledges support by the Ram\'on y Cajal
programme by the Ministerio de Ciencia e Innovaci\'on of the
government of Spain.
We also thank the referee for his constructive comments.
}

\end{document}